\documentclass[epj]{svjour}
\usepackage{graphics}
\usepackage{amsmath,amssymb}
\newcommand{\bc}{\begin{center}}
\newcommand{\ec}{\end{center}}
\newcommand{\be}{\begin{equation}}
\newcommand{\ee}{\end{equation}}
\newcommand{\ba}{\begin{array}}
\newcommand{\ea}{\end{array}}
\newcommand{\bea}{\begin{eqnarray}}
\newcommand{\eea}{\end{eqnarray}}
\newcommand{\bt}{\begin{tabular}}
\newcommand{\et}{\end{tabular}}

\newcommand{\bsl}{\boldsymbol}

\begin{document}
\title{Accuracy of Auxiliary Field Approach for Baryons}
\author{I.M.Narodetskii\inst{1} \and C.Semay\inst{2}\and A.I.Veselov\inst{1}
\thanks{\emph{This work was supported by RFBR grants
06-02-17120, 08-02-00657, and 08-02-00677. C. Semay  thanks the
F.R.S.-FNRS for financial support.
}
}%
}                     
\offprints{I.M.Narodetskii}          
\institute{Institute of Theoretical and Experimental Physics,
117218 Moscow, Russia \and Groupe de Physique Nucl\'{e}aire
Th\'{e}orique, Universit\'{e} de Mons-Hainaut, Place du Parc 20,
BE-7000 Mons, Belgium. }
\date{Received: date / Revised version: date}
%
\abstract{ We provide a check of the accuracy of the auxiliary
field formalism used to derive the Effective Hamiltonian for
baryons in the Field Correlator Method.  To this end we compare
the solutions for the Effective Hamiltonian with those obtained
from the solution of the spinless Salpeter equation. Comparing
these results gives a first estimate of the systematic uncertainty
due to the use of the auxiliary field formalism for baryons.
} 
\maketitle
\section{Introduction}
\label{intro} The advent of new ideas concerning quark-quark
forces in QCD has led to a revival of interest in baryon
spectroscopy. Various versions of the constituent quark model
\cite{QM} reproduce the octet and decuplet ground states but have
very different and even contradictory predictions on the spectrum
of excited states. It is therefore very important to develop model
independent methods that are directly connected to the QCD
Lagrangian and can help in alternatively understanding baryon
spectroscopy.

One of such approaches is based on the Field Correlator Method
(FCM) in QCD \cite{DS}. FCM provides a promising formulation of
the nonperturbative QCD that gives additional support for the
quark model assumptions. The application of this method for light
mesons, heavy quarkonia, heavy-light mesons and light and heavy
baryons can be found in Refs. \cite{BBS}. The key ingredient of
the FCM is the use of the auxiliary fields (AF) initially
introduced in order to get rid of the square roots appearing in
the relativistic Hamiltonian~\footnote{Historically the AF
formalism was first introduced in  \cite{P} to treat the
kinematics of the relativistic spinless particles. For a brief
review of the AF formalism relevant to the problem considered in
this paper see Sec. II of \cite{KNS}.}. Using the AF formalism
allows one to write a simple local form of the Effective
Hamiltonian (EH) for the three quark system \cite{S2003}, which
comprises both confinement and relativistic effects and contains
only universal parameters: the string tension $\sigma$, the strong
coupling constant $\alpha_s$, and the bare (current) quark masses
$m_i$. The EH has the form
\begin{equation}
\label{eq:H} H=\sum\limits_{i=1}^3\left(\frac {m_{i}^2}{2\,\mu_i}+
\frac{\mu_i}{2}\right)+H_0+V.
\end{equation}
In Eq. (\ref{eq:H}), $H_0$ is the non-relativistic kinetic energy
operator for masses $\mu_i$, $V$ is the sum of the string
potential $V_{Y}({\bf r}_1,\,{\bf r}_2,\,{\bf r}_3)$ and a Coulomb
interaction term $V_{\rm Coulomb}$ arising from the one-gluon
exchange. The string potential is \be V_Y({\bf r}_1,\,{\bf
r}_2,\,{\bf r}_3)\,=\,\sigma\,r_{min},\ee where $r_{min}$ is the
minimal string length corresponding to the Y-shaped configuration.
Finally the $\mu_i$ are the operator AF that have to be determined
from the variational principle.

Note that the the sum of the mass term and $H_0$ in Eq.
(\ref{eq:H}) can be conveniently written as
\begin{equation}
\sum\limits_{i=1}^3\left(\frac
{m_{i}^2}{2\,\mu_i}\,+\,\frac{\mu_l}{2}\right)\,+\,H_0\,=\,
\sum\limits_{i=1}^3\left(\frac {{\bf
p}_i^2\,+\,m_{i}^2}{2\,\mu_i}\,+\,\frac{\mu_i}{2}\right).\end{equation}
After taking the extremum of this expression in $\mu_i$ one ends
with the standard relativistic kinetic energy operator
$\sum\limits_{i=1}^3\,\sqrt{{\bf p}_i^2+\,m_i^2\,}$.

In this paper we use an approximate approach to consider the AF
formalism first suggested in \cite{DKS}. The AF are treated as
c-number variational parameters. In this approach one replaces the
operators $\mu_i(\tau)$ depending on time parameter $\tau$ by the
c-numbers $\mu_i$ independent of $\tau$. The eigenvalue problem is
solved for each set of $\mu_i$; then one has to minimize $\langle
H\rangle$ with respect to $\mu_i$. Such an approach allows for a
very transparent interpretation of AF: starting from bare quark
masses $m_i$, we naturally arrive at the dynamical masses $\mu_i$
that appear due to the interaction and can be treated as the
dynamical masses of constituent quarks.

An obvious disadvantage of the AF approach is that, as a
variational method, it provides only an upper bound to the mass
spectrum. So far the accuracy of this approximate solution for
relativistic systems has been checked numerically only for mesons
\cite{KNS,sema04}. The principle objective of this work is to test
the AF method for baryons. We implement the AF method to calculate
the baryon masses and then perform similar calculations using the
relativistic Hamiltonian \be H\,=\,\sum\limits_{i=1}^3\,\sqrt{{\bf
p}_i^2\,+\,m_i^2}\,+V. \label{eq:sse} \ee Although being formally
simpler the Hamiltonian (\ref{eq:H}) is equivalent to
(\ref{eq:sse}) up to the elimination of the AF (see e.g.
Ref.~\cite{BM}). We refer to an eigenvalue equation with
Hamiltonian (\ref{eq:sse}) as the spinless Salpeter equation
(SSE). In QCD, it arises from the Bethe-Salpeter equation
replacing the interaction by the instantaneous potential $V$ and
considering a limited Fock space containing $qqq$ states only.

In this paper, we study the confinement plus Coulomb energies for
the ground $S$-wave and orbitally excited $P$-wave states of
$nnn$, $nns$ and $ssn$ baryons~\footnote{Here and below the symbol
$n$ stands for the light quarks $u$ or $d$.} and disregard the
spin dependent forces, which are not relevant for our
consideration.

The baryon masses in the AF approach are calculated using the
hyperspherical method, while those in the SSE are calculated
variationally. The numerical algorithm to solve the three-body
problem variationally is based on an expansion of the wave
function in terms of harmonic oscillator functions with different
sizes \cite{nunb77}. The details of technical aspects can be found
elsewhere \cite{silv01}. It was proved to give results of good
accuracy if the expansion is pushed sufficiently far (let say up
to 16-20 quanta). Moreover it can deal easily either with a
non-relativistic or relativistic expression for the kinetic energy
operator.

We find an accuracy of the AF method for hyperons to be about 6
$\%$ at worst, which is quite reasonable to justify application of
the AF formalism.

The paper is organized as follows. In Sec. 2, we briefly review
the EH method. The application of this method for the baryons was
described in detail elsewhere \cite{NT,DNV}.
 Here we
give only a brief summary important for our particular
calculation. In Sec. 3, we discuss the hyperspherical approach,
which is a very effective numerical tool to solve this
Hamiltonian. In Sec.4, we provide a few numerical examples
illustrating the accuracy of the hyperspherical solutions. In Sec.
5, predictions of the AF method are compared with those obtained
from the solution of the spinless Salpeter equation (SSE). Section
6 contains our conclusions.
\section{The baryon masses in  the AF method and SSE}
The baryon mass in the FCM is given by
\begin{equation}
\label{M_B} M_B^{AF}\,=\,M_0^{AF}\,+\,C^{AF},\end{equation}
\begin{equation}
\label{eq:M_B0}M_0^{AF}\,=\,\sum\limits_{i=1}^3\left(\frac
{m_{i}^2}{2\mu_i\,}+
\,\frac{\mu_i}{2}\right)\,+\,E_0(\mu_i)\end{equation} where
$E_0(\mu_i)$ is an eigenvalue of the Shr\"{o}dinger operator $H_0
+V$, the constant AF $\mu_i$ are defined from the minimum
condition \be\label{eq:mc}
\frac{\partial\,M_0^{AF}(m_i,\mu_i)}{\partial\,\mu_i}\,
=\,0,\end{equation} and $C^{AF}$ is the quark self-energy
correction which is created by the color magnetic moment of a
quark propagating through the vacuum background field
\cite{S2001}. This correction, which can be added perturbatively,
adds an overall negative constant to the hadron masses:
\begin{equation} \label{self_energy}
C^{AF}\,=\,-\frac{2\sigma}{\pi}\,\sum\limits_i\frac{\eta(t_i)}{\mu_i},\,\,\,\,\,t_i\,=\,m_i/T_g,\end{equation}
where $1/T_g$ is the gluonic correlation length. In what follows
we use $T_g\,=\, 1$ GeV.

The function $\eta(t)$ is defined as
\be \eta(t)= t\int^\infty_0 z^2\, K_1(tz)\,
e^{-z}\,dz,\label{eq:eta} \ee where $K_1$ is the McDonald
function. A straightforward calculation yields \cite{S2001}

\bea
\eta(t)&=&\frac{1+2t^2}{(1-t^2)^2}-
\frac{3t^2}{(1-t^2)^{5/2}}\, \ln\,
\frac{1+\sqrt{1-t^2}}{t},\quad t<1, \nonumber\\
&=&\frac{1+2t^2}{(1-t^2)^2}-
\frac{3t^2}{(t^2-1)^{5/2}}\,\arctan\,
(\sqrt{t^2-1}),\quad t>1
\eea
Note that
$\eta(0)\,=\,1$ and $\eta(t)\,\sim\,2/t^2$ as $t\,\to\,\infty$.

The baryon mass in the SSE approach is given by \be\label{M_SSE}
M_B^{SSE}=M_0^{SSE}\,+\,C^{SSE},\ee where $M_0^{SSE}$ is an
eigenvalue of the relativistic Hamiltonian (\ref{eq:sse}) and the
$C^{SSE}$ are given by (\ref{self_energy}) with the obvious
substitution $\mu_i\,\to\,\omega_i$, where \be
\omega_i\,=\,\langle\,\sqrt{{\bf
p}_i^2\,+\,m_i^2}\,\rangle\label{eq:omega_i}\ee are the average
kinetic energies of the current quarks.

 We will not perform a
systematic study in order to determine the best set of parameters
to fit the baryon spectra. Instead, in what follows we employ some
typical values of the string tension $\sigma$ and the strong
coupling constant $\alpha_s$, which have been used for the
description of the ground state baryons \cite{NT}: $\sigma\,=\,$
0.15 GeV$^2$ and $\alpha_s\,=\,$ 0.39. In our calculations we use
the values of the current light quark masses,
$m_u\,=\,m_d\,=\,9\,$ MeV, and $m_s\,=\,175$ MeV. As in Ref.
\cite{NT} we neglect the spin dependent potentials responsible for
the fine and hyperfine splittings of baryon states.

Our aim is to compare the baryon masses given by Eqs. (\ref{M_B})
and (\ref{M_SSE}). To this end we first solve the non-relativistic
Schr\"odinger equation with the confining and Coulomb interactions
to determine the constituent quark masses $\mu_i$ and the baryon
masses $M_{B}^{AF}$. Efficient methods to deal with the Y-shape
interaction rely either on Monte-Carlo
algorithms~\cite{carl83,sart85} or the hyperspherical
method~\cite{fabr91}. We use the latter approach.

\section{Outline of the hyperspherical formalism.}
In this section, we briefly review the hyperspherical method,
which we use to calculate the masses of the ground and excited
hyperon states.

The baryon wave function depends on the three-body Jacobi
coordinates \bea\label{eq:Jacobi}
&&\bsl{\rho}_{ij}=\sqrt{\frac{\mu_{ij}}{\mu_0}}\,(\bsl{r}_i-\bsl{r}_j),\nonumber\\
&&\bsl{\lambda}_{ij}=\sqrt{\frac{\mu_{ij,\,k}}{\mu_0}}
\left(\frac{\mu_i\bsl{r}_i+\mu_j\bsl{r}_j}{\mu_i+\mu_j}-\bsl{r}_k\right),\eea
($i,j,k$ cyclic), where $\mu_{ij}$ and $\mu_{ij,k}$ are the
appropriate reduced masses:
\begin{equation}\mu_{ij}=\frac{\mu_i\mu_j}{\mu_i\,+\,\mu_j},~~~~
\mu_{ij,\,k}=\frac{(\mu_i\,+\,\mu_j)\mu_k}{\mu_i\,+\,\mu_j\,+\,\mu_k},\end{equation}
and $\mu_0$ is an arbitrary parameter with the dimension of mass,
which drops out in the final expressions. There are three
equivalent ways of introducing the Jacobi coordinates, which are
related to each other by linear transformations with the Jacobian
equal to unity. In what follows we omit the indices $i$ and $j$.

In terms of the Jacobi coordinates the kinetic energy operator
$H_0$ in (\ref{eq:H}) is written as \bea \label{H_0_jacobi} &&H_0=
-\frac{1}{2\mu_0} \left(\frac{\partial^2}{\partial\bsl{\rho}^2}
+\frac{\partial^2}{\partial\bsl{\lambda}^2}\right)\,=\nonumber\\
&&=\,-\,\,\frac{1}{2\mu_0}\left( \frac{\partial^2}{\partial
R^2}+\frac{5}{R}\frac{\partial}{\partial R}+
\frac{\bsl{L}^2(\Omega)}{R^2}\right), \eea where $R$ is the
six-dimensional hyperradius that is invariant under quark
permutations, \bea
&&R^2\,=\,\bsl{\rho}^2+\bsl{\lambda}^2,\nonumber\\&&
\rho\,=\,R\,\sin\theta,\,\,\,\,\,
\lambda\,=\,R\,\cos\theta,\,\,\,\, 0 \le \theta \le \pi/2,\eea
$\Omega$ denotes five residuary angular coordinates, and
$\bsl{L}^2(\Omega)$ is an angular operator
 \be{\bf
L}^2\,=\,\frac{\partial^2}{\partial
\theta^2}\,+\,4\cot\theta\,\frac{\partial}{\partial
\theta}-\frac{{\bf l}_{\rho}^2}{\sin^2\theta}\,-\,\frac{{\bf
l}_{\lambda}^2}{\cos^2\theta},\ee
whose eigenfunctions (the hyperspherical harmonics) satisfy
\begin{equation}
\label{eq: eigenfunctions} {\bf L}^2(\Omega)\,Y_{[K]}(\theta,{\bf
n}_{\rho},{\bf n}_{\lambda})\,=\,-K(K+4)Y_{[K]}(\theta,{\bf
n}_{\rho},{\bf n}_{\lambda}),
\end{equation}
with $K$ being the grand orbital momentum.

The wave function $\psi(\bsl{\rho},\bsl{\lambda})$ is written in a
symbolical shorthand as
\be\psi(\bsl{\rho},\bsl{\lambda})=\sum\limits_{[K]}\psi_{[K]}(R)Y_{[K]}(\Omega),\label{eq:ss}\ee
where the set $[K]$ is defined by  the orbital momentum of the
state and the symmetry properties.

We truncate this set using the approximation $K\,=\,K_{\rm min}$.
We comment on the accuracy of this approximation latter on. Our
task is then extremely simple in principle: we have to choose a
zero-order wave function corresponding to the minimal $K$ for a
given $L$ ($K_{\rm min}\,=\,0 $ for $L\,=\,0$ and $K_{\rm
min}\,=\,1 $ for $L\,=\,1$). The corresponding hyperspherical
harmonics are \bea &&
Y_0\,=\,\sqrt{\frac{1}{\pi^3}}\,,\,\,\,\,K\,=\,0,\nonumber\\&&
\bsl{Y}_{\rho}\,=\,\sqrt{\frac{6}{\pi^3}}\,\frac{\bsl{\rho}}{R}\,,\,\,\,\,\,\,\,
\bsl{Y}_{\lambda}\,=\,\sqrt{\frac{6}{\pi^3}}\,\frac{\bsl{\lambda}}{R}\,,\,\,\,\,K\,=\,1.\eea

For $nns$ baryons we use the basis in which the strange quark is
singled out as quark $3$ but in which the non-strange quarks are
still antisymmetrized. In the same way, for the $ssn$ baryon we
use the basis in which the non strange quark is singled out as
quark $3$.  The $nns$ basis states diagonalize the confinement
problem with eigenfunctions that correspond to separate
excitations of the non-strange and strange quarks (${\rho}$\,- and
${\lambda}$\, excitations, respectively). In particular,
excitation of the $\bsl{\lambda}$ variable unlike excitation in
$\bsl{\rho}$ involves the excitation of the ``odd'' quark ($s$ for
$nns$ or $n$ for $ssn$). The nonsymmetrized $uds$ and $ssq$ bases
usually provide a much simplified picture of the states. The
physical P-wave states are neither pure SU(3) states nor pure
$\rho$ or $\lambda$ excitations but  linear combinations of all
states with a given $J$. Most physical states are, however, closer
to pure $\rho$ or $\lambda$ states than to pure SU(3) states
\cite{IK}. Note that for the $nnn$ baryons, the ${\bsl\rho}$ and
${\bsl\lambda}$ excitation energies are degenerate.

Introducing the reduced function $u_{\gamma}(R)$
\be\Psi_{\gamma}(R,\Omega)\,=\,\frac{u_{\gamma}(R)}{R^{5/2}}\cdot{
Y}_{\nu}(\Omega), \ee where $\gamma\,=\,0$~ for $L\,=\,0$,~
$\gamma\,=\,\rho,\,\lambda$ for $L\,=\,1$~\footnote{In what
follows, for ease of notation  we will drop the magnetic quantum
numbers of the vector spherical harmonics.}, the new variable
\bea&&
x\,=\,\sqrt{\mu_0}\,R\,=\nonumber\\
&&\left(\sum_i\,\frac{\mu_1\,\mu_2}{M}\,r_{12}^2\,+\,\frac{\mu_2\,\mu_3}{M}\,r_{23}^2\,+\,
\frac{\mu_3\,\mu_1}{M}\,r_{31}^2\right)^{1/2},\eea and averaging
the interaction $V=V_Y+ V_{C}$ over the six-dimensional sphere
$\Omega$ with the weight $|Y_{\gamma}|^2$, one obtains the
one-dimensional Schr\"odinger equation for $u_{\gamma}(x)$
 \bea\label{eq:se}&&\frac{d^2
u_{\gamma}(x)}{dx^2}\,+\nonumber\\&&2\left(E_0\,-\,\frac{(K+\frac{3}{2})(K+\frac{5}{2})}{2\,x^2}\,-
\,V_{\gamma}(x)\right)u_{\gamma}(x)\,=\,0,\eea where $
V_{\gamma}(x)\,=\,V_{\rm Y}^{\,\gamma}(x)\,+\,V_{\rm
Coulomb}^{\,\gamma}(x),$ \bea\label{string} &&V_{\rm
Y}^{\,\gamma}(x)\,=\,\int \,|Y_{\gamma}\,(\theta,\chi)|^2\,V_{\rm
Y}({\bf r}_1,\,{\bf r}_2,\,{\bf r}_3)\,d\Omega\,=\nonumber\\&&
\sigma\, b_{\nu}\,\frac{x}{\sqrt{\mu_0}},\eea
and\bea\label{Coulomb}&& V_{\rm
Coulomb}^{\,\gamma}(x)\,=\,-\,\frac{2}{3}\,\alpha_s\,\int
\,|Y_{\gamma}\,(\theta,\chi)|^2\,\sum_{i\,<\,j}\,\frac{1}{r_{ij}}\,\,\,d\Omega\,=\nonumber\\&&
-\,\frac{2}{3}\,\alpha_s\,\frac{a_{\gamma}}{x}\,\sqrt{\mu_0}.\eea
  In what follows we denote
\be\label{eq:definition}\mu_1\,=\,\mu_2\,=\mu,\,\,\,\,\, \,
\mu_3\,=\,\kappa\,\mu.\ee Then the straightforward analytical
calculation of the integrals in (\ref{Coulomb}) yields \be
a_{0}\,\sqrt{\mu_0}\,=\,\frac{16}{3\,\pi}\left(\,\sqrt{2}\,+\,2\,\sqrt{\frac{\kappa}{1\,+\,\kappa}}\,
\right)\,\sqrt{\mu}, \ee
\be
a_{\rho}\,\sqrt{\mu_0}\,=\,\frac{32}{15\,\pi}\left(\,\sqrt{2}\,+\,\sqrt{\frac{\kappa}{1\,+\,\kappa}}\,\frac{5\kappa+6}{1\,+\,\kappa}
\right)\,\sqrt{\mu},\ee
\be
a_{\lambda}\,\sqrt{\mu_0}\,=\,\frac{32}{15\,\pi}\left(\,\frac{3}{\sqrt{2}}\,+\,\sqrt{\frac{\kappa}{1\,+\,\kappa}}\,
\frac{4\,+\,5\kappa}{1\,+\,\kappa}\right)\,\sqrt{\mu}.\ee
 For $\kappa\,=\,1$ (the $nnn$ system) $a_{\rho}\,=\,a_{\lambda}$.
 The corresponding
expressions for $b_{\gamma}$ are more complicated (see, e.g., the
appendix of Ref. \cite{DNV}).

\section{Accuracy of the hyperspherical approximation}

A few words concerning the accuracy of the approximation $K=K_{\rm
min}$ are in order. An illustration of the accuracy of the
hyperspherical approximation $K=K_{\rm min}$ is given by the
results presented in Table \ref{tab:E_0_comparison}. This Table
compares the eigenvalues $E_0$ in Eq. (\ref{eq:M_B0}) for the
$nnn$, $nns$ and $ssn$ systems obtained using the variational
method and those calculated from Eq. (\ref{eq:se}) with
$K\,=\,K_{\rm min}$~\footnote{Recall that, as was stated in Sec.
3, the ${\bsl\rho}$ and ${\bsl\lambda}$ excitation energies for
the $nnn$ baryon are degenerate.}. In all cases the dynamical
masses $\mu_i$ are the same as were found from the minimum
condition (\ref{eq:mc}) for the Y-shaped string potential
\cite{DNV}. For technical reasons the variational calculations
have been performed not for the genuine string junction potential
but for its approximation by a sum of the one- and two-body
confining potentials \cite{SSNV} \be\label{eq:V_M} V_{\text
M}\,=\,\frac{1}{2}\,(V_{\Delta}\,+\,V_{\text{CM}}),\ee where
$V_{\Delta}$ with the sum of the two-body confining potentials is
\be \label{eq:Delta}
V_{\Delta}\,=\,\sigma\,\frac{1}{2}\,\sum_{i<j}\,r_{ij}\,=\,\sigma\,\frac{1}{2}\,\sqrt{\mu_0}\,\,\sum_{i\,<\,j}\,\frac{|{\bsl
\rho}_{ij}|}{\sqrt{\mu_{ij}}}, \ee and $V_{\text{C}}$ is the sum
of one-body center-of-mass string potentials: \bea \label{eq:cms}
V_{\text{CM}}&=&\sigma\,\sum\limits_i\,|{\bsl r}_i\,-\,{\bsl
R}_{cm}| \\ \nonumber &=&\sigma\,\sqrt{\mu_0}\,\sum_{(i,j,k)}
\frac{1}{\mu_k}\,\sqrt{\mu_{ij,\,k}}\, |\bsl{\lambda}_{ij}|, \eea
($i,j,k$ cyclic), where ${\bsl R}_{cm}$ is the center-of-mass
coordinate. Table \ref{tab:E_0_comparison} also compares
eigenvalues $E_0^Y$ for the genuine string  potential $V_Y$ with
those for the confining potentials $V_{\Delta}$, $V_{\text CM}$
and $V_{\text M}$ with the same string tension. The confining
potential $V_{\text CM}$ overestimates the eigenvalues of the
genuine string junction $E_0^Y$ while the potential $V_{\Delta}$
underestimates the $E_0^Y$, i.e.
$E_0^{\Delta}\,<\,E_0^{Y}\,<E_0^{\text CM}$ (compare columns $6$,
$7$ and $8$ of Table \ref{tab:E_0_comparison}). The values of the
two columns $7$ and $8$ are in reasonable agreement with the
reference results of column $6$. In line with expectations
\cite{SSNV}, the eigenvalues for the genuine string junction
change little if we use $V_{\text M}$ instead of $V_{\text Y}$.
Simulation of the genuine string junction potential by a sum of
the two-body confining potentials (\ref{eq:V_M}) (column $9$) is a
good approximation in all cases: using $V_{\text M}$ results in a
$\sim\,$ 20 MeV or $1\,-\,2\,\%$ downwards shift of $E_0$ for all
states (compare columns $6$ and $9$). Let us note that
Hamiltonian~\ref{eq:H} with potential $V_M$ gives eigenvalues wich
are, to some MeV, the arithmetic mean of the eigenvalues with
potential $V_\Delta$ and $V_{CM}$. So the contributions of
$V_\Delta$ and $V_{CM}$ to $V_M$ are nearly evenly distributed.

The last column $10$ contains the eigenvalues $E_{0~\text
var}^{M}$ calculated using the variational method briefly
described in Sect. 1. Comparing the column $9$ and $10$ of Table
\ref{tab:E_0_comparison} we conclude that the hyperspherical and
variational results are close enough  to validate the
approximation $K=K_{\rm min}$.


\section{Comparison of the AF and SSE results}

Table \ref{tab:comparison} compares the baryon masses computed
using the AF and SSE formalisms. In this Table we list the masses
of the $nnn$, $nns$ and $ssn$ states with $L\,=\,$ 0,1. The
entries labeled $AF$ have been calculated from Eq. (\ref{eq:se})
with $K\,=\,K_{\rm min}$, while the entries labeled $SSE$ have
been calculated using the variational method for the relativistic
Hamiltonian (\ref{eq:sse}). In both cases,
we approximate the Y shaped string potential by the expression
(\ref{eq:V_M}). As was mentioned in the Introduction the
comparison of the AF results with those evaluated from the
solution of SSE has been performed only for the ${\overline q}q$
mesons with the conclusion that the variational AF method gives a
systematic overestimation of order 5-7 $\%$ \cite{KNS,sema04}. Our
calculations show the similar results: the relative deviation
\be\varepsilon\,=\,\frac{M_B^{AF}\,-\,
M_B^{SSE}}{M_B^{SSE}}\label{eq:varepsilon}\ee is positive and  for
most considered states does not exceed 6$\%$~\footnote{An obvious
exception is the $nnn$ state with $L\,=\,0$ for which
$\varepsilon$ reaches 14$\%$.}. The accuracy of the AF approach
does not seem to be very sensitive to the bare light-quark masses.
The quantum numbers of states have a stronger influence on the
accuracy. In particular, $\varepsilon$ for the $L\,=\,1$ states
are uniformly smaller than those for the $L\,=\,0$ states.
Curiously, the self-energy corrections $C^{MFC}$ and $M^{SSE}$
agree even with better accuracy (typically within 5$\%$ or even
better) in spite of the fact that the difference $\mu_i$ and
$\omega_i$ in some cases (e.g. for the $\lambda$ excitation in the
$ssn$) comprises 30$\%$. As for the excitation energies,
$\Delta\,=\,M_B(L=1)\,-\,M_B(L=0)$ evaluated using the AF and SSE
methods, they practically coincide for the $ssn$ baryons and
differ no more than $\sim 30$ MeV for the $nns$ baryons. Taking
into consideration that we neglect the spin interactions the
baryon energies calculated using SSE agree reasonably with the
data \cite{PDG06}. For instance, for $L\,=\,0$ we get
$\frac{1}{2}\,(N\,+\,\Delta)_{\rm theory}\,=\,$ 1062 MeV versus
$\frac{1}{2}\,(N\,+\,\Delta)_{\rm exp}\,=\,$ 1085 MeV and
$\frac{1}{4}(\Lambda\,+\,\Sigma\,+\,2\,\Sigma^*)_{\rm
theory}\,=\,$ 1220 MeV versus
$\frac{1}{4}(\Lambda\,+\,\Sigma\,+\,2\,\Sigma^*)_{\rm exp}\,=\,$
1267 MeV. A similar correspondence exists for the other states
considered in this work.

\section{Conclusions}
In this paper we have tested the quality of our previous study of
the masses of the S- and P- baryon states obtained in the FCM with
the use of the AF formalism. To this end we have compared  the AF
results with those obtained from the solution of the SSE with the
same interaction. The main purpose was to check whether the
results obtained within these two methods are similar. We have
found that they agree within $\sim$ 100 MeV for the absolute
values of masses and with much better accuracy for the excitation
energies. Thereby our study supports the AF basic assumptions by
the compatibility of its mass predictions with the masses derived
from the SSE. Moreover, this comparative study gives  better
insight into the quark model results, where the constituent masses
encode the QCD dynamics.


\begin{table*}[t]
\caption{Comparison of the eigenvalues $E_0$ of the Hamiltonian
$H_0\,+\,V$ in Eq. (\ref{eq:H}) for the baryon ground states and
the $\rho$ and $\lambda$ excitations obtained from the
hyperspherical solution of Eq. (\ref{eq:se}) ($E_0^Y$) and
variational solution ($E_{0~\it var}^{\text M}$). See the text for
further explanation.} \label{tab:E_0_comparison} \centering
\begin{tabular}{cccccccccc} \hline\hline\\
Baryon&$L$ &Excitation&$\mu_1$ & $\mu_3$& $E_0^Y$&
$E_0^{\text CM}$&$E_0^{\Delta}$&$E_0^{\text M}$&$E_{0~\it var}^{\text M}$\\
\\ \hline\\
$nnn$&0&&408&408&1318&1366&1230&1299&1297\\
&1&$\rho,\,\lambda$&457&457&1638&1697&1532&1615&1612\\ \\
\hline \\
$nns$&$0$&&414&453&1291&1339&1204&1272&1271\\
&$1$&$\rho$&482&459&1611&1670&1506&1589&1587\\
&$1$ &$\lambda$&441 &534 &1614 & 1676&1508&1593&1591\\
\\\hline
\\
$ ssn$&$0$&&458&419&1266&1313&1181&1248&1248\\
 & $1$& $\rho$ & 520 & 424 & 1592&1653&1487&1571&1569 \\
&$1$& $\lambda$&483 &506&1588&1646&1485&1567&1566\\

\\

\hline\hline
\end{tabular}

\end{table*}
~~\\[2cm]

\begin{table*}[t] \caption{Comparison of baryon masses calculated using the AF
approach and SSE.  The symbol~ $\nu_i$ denotes either the
constituent quark masses $\mu_i$ or the average kinetic energies
of the current quarks $\omega_i$. Shown are the masses $M_0^{AF}$
and $M_0^{SSE}$ without the self-energy corrections, the
self-energy corrections $C^{AF}$ and $C^{SSE}$, $M\,=\,M_0\,+\,C$
(all in units of MeV), and the relative error $\varepsilon$
defined by Eq. (\ref{eq:varepsilon}).
}\centering
\begin{tabular}{cccccccccc} \hline\hline\\
Baryon &L&Excitation&Method& $\nu_1\,=\,\nu_2$ & $\nu_3$& $M_0$&C&M&$\varepsilon(\%)$\\
\\ \hline\\
$nnn$&0&&AF&408&408&1911&$-\,702$&1209&13.8\\
&&&SSE&394&394&1788&$-\,726$&1062&\\ \\
 $nns$&0&  & AF& 414& 453& 1946&$-\,648$&1298&6.4\\
 && & SSE&396 & 484 &   1877&$-\,657$&1220&\\ \\
 $ssn$&0&&AF&458&419&1182&$-\,598$&1384&6.3\\
 &&&SSE&404&465&1904&$-\,602$&1302&\\ \\
 \hline\\
 $nnn$&1&$\rho,\,\lambda$&AF&457&457&2301&-627&1674&9.1\\
 &&$\rho,\,\lambda$&SSE&440&440&2186&-651&1534\\ \\
 $nns$&1& $\rho$  & AF&482 & 459 & 2356&$-\,581$&1751&6.0\\
  & &$\rho$  & SSE&464 & 465 &  2245& $-\,594$&1652&\\
& &$\lambda$&AF&441 &534 & 2330&$-\,592$&1738&1.5\\
&&$\lambda$&SSE&415 &592& 2315& $-\,603$&1712&\\ \\
$ssn$ &1&  $\rho$  & AF&520 & 424 & 2362&$-\,552$&1810&4.1\\
&&$\rho$&SSE&478&530&2302&$-\,564$&1738&\\
& &$\lambda$ &AF&483 &506&2367&$-\,540$&1827&4.4\\
&&$\lambda$&SSE&503&391&2295&$-\,545$&1750&\\

\\
\hline\\
\end{tabular}
\\

\label{tab:comparison}
\end{table*}

\end{document}